\documentclass[aps,prl,twocolumn,floats,showpacs]{revtex4}
\usepackage{graphicx}
\usepackage{color}
\usepackage{bm}

\newcommand{\eexp}{\mbox{e}^}
\begin{document}
\title{Shot noise in a Majorana fermion chain }
\author{  Anatoly Golub and  Baruch Horovitz}
\affiliation{ Department of Physics, Ben-Gurion University, Beer Sheva 84105 Israel\\   }
\begin{abstract}
We calculate the shot noise power in a junction of  a network of Majorana bound states (MBS) with a normal metal.
These Majorana bound states are on the border of alternative ferromagnetic and superconducting regions at the quantum spin Hall insulator edge. We analyze different realization of MBS networks including a few isolated ones or in a chain allowing for the limit of weak and strong tunneling. The conductance, as well as the shot noise are considerably stronger than those of a weakly coupled normal-superconduting junction, being a hallmark of the MBS. We find that the Fano factor is quantized $F=2$ when one MBS member of a pair is coupled to the lead, however if both MBS members are coupled $F$ is non-integer.
\end{abstract}
\maketitle
{\it Introduction}: Recently discovered topological materials in combination with a superconductor can host Majorana fermions\cite{fu,sato,oreg,das,alicea,potter,qi,been}. Interest in such quasiparticles is due to the possibility  of non-Abelian statistics which they satisfy \cite{nayak}. The most accessible case are Majorana Bound States (MBS) at the 1-dimensional edge of a quantum spin Hall insulator \cite{kitaev,fu2}. Two Majorana fermions can form a usual Dirac fermion. However, to detect a state of Majorana fermion which has no definite charge, requires non-local measurements. It was suggested \cite{law,linder} that a tunneling probe can detect the presence of MBS. While the tunneling probe is a local measurement it can detect interference effects between various MBS.  Recently an array of alternating ferromagnetic and superconducting regions at the QSH edge were considered \cite{refael,flens}. This array may appear due to local density fluctuations at a random FM/SC boundaries that host MBS. The tunneling characteristics (conductance and tunneling current) were calculated for a network of coupled Majorana fermions. Both isolated \cite{flens} and random chains \cite{refael} cases were analyzed.  Here we show how Majorana fermions can be detected by shot noise measurements. For this we calculate the noise power for different networks of MBS and present a general formulation of this problem based on Keldysh technique.

{\it The Hamiltonian}:  We consider the tunneling between the  normal metal lead and a realization of 1d Majorana chain \cite{fu2,alicea2,refael,flens}. The coupled Majorana state network is described by the Hamiltonian
\begin{eqnarray}
H_M&=&\frac{i}{2}\sum_{i,j}t_{ij}\gamma_i \gamma_j
 \label{HM}
\end{eqnarray}
where $\gamma_i$ are operators for the $i$-th Majorana fermion, satisfying $\gamma_i= \gamma_i^{\dagger}$, $\gamma_i^2=1$, and $t_{ij}$ are elements of an antisymmetric matrix ${\hat t}$ for the coupling between the $i,j$ MBS. Disorder may be introduced as random nearest neighbor coupling. The $\gamma_i$ can be written as a Bogolyubov transformation for the quasiparticles in the superconductor composed of creation $\Psi_{\sigma}^{\dagger}(x)$ and annihilation $\Psi_{\sigma}(x)$  operators of electrons at position $x$ and spin $\sigma$. The coefficients $f_{\sigma,i}(x)$ in this transformation are the eigenvalues of the Bogoloyubov de Gennes Hamiltonian with zero energy. The general form is
\begin{equation}\label{ma}
\gamma_i=\Sigma_\sigma\int dx (f_{\sigma,i}(x)\Psi_{\sigma}(x)+f_{\sigma,i}^*(x)\Psi_{\sigma}^{\dagger}(x))
\end{equation}

The tunneling between the normal metal electrode and the superconductor is given by ${\cal H}_T=\sum_{k\sigma}\int dx[t_k(x)c_{k\sigma}^{\dagger}\Psi_{\sigma}(x)+h.c.]$. Below we consider the energy gap $\Delta$ of superconductor as the biggest energy scale in the problem. Then for small  applied voltage $eV<\Delta$  only zero-energy Majorana operator projection of total quasiparticle operator in superconductor   is important \cite{refael,fu3,flens}. Thus the tunneling Hamiltonian and the current operator in  Nambu space become
\begin{eqnarray}
H_T&=&\frac{1}{2}\sum_{i}\bar{c}(0)\tau_z \hat{V}_i \gamma_i +h.c
 \label{HT}\\
 I&=&\frac{ie}{2}\sum_{i}\bar{c}(0)\hat{V}_i \gamma_i +h.c \equiv ieJ
 \label{I}
\end{eqnarray}
where
$c(0)=(c_{\uparrow},c_{\downarrow},c_{\downarrow}^{\dagger},-c_{\uparrow}^{\dagger})^T$ corresponds to the normal lead electron operator at $x=0$ 
 The coupling  between the normal lead and Majorana states is given by a vector in Nambu space $\hat{V}_i=(\bar{f}_{\uparrow,i},\bar{f}_{\downarrow,i},\bar{f}_{\downarrow,i}^{*},-\bar{f}_{\uparrow,i}^{*})^T$
 where $\bar{f}_{\sigma,i}=\int dx f_{\sigma,i}(x)t_{k}(x)$ describe the interaction of MBS with the lead electrons and we assumed a weak momentum dependence of matrix element $t_{k}$ for tunneling to the lead. The Pauli $\tau$-matrices act on ($c_{\sigma}$; $\bar{f}_{\sigma,i}$) and ($c_{\sigma}^{\dagger}$; $\bar{f}_{\downarrow,i}^{*}$) blocks. Total hamiltonian in addition to parts (\ref{HM},\ref{HT})
includes the lead contribution which we take at voltage bias $V$. We write below the action for lead in the rotated Keldysh basis
\begin{eqnarray}
  S_{lead} &=& \int dt\Sigma_k \bar{\hat{c}}_k g_k^{-1}\hat{c}_k
\end{eqnarray}
where $g_k^{-1}$ is the inverse Green function for electrons in the lead. For further use we introduce the Green's function (GF)  of the lead integrated on momentum
$\bar{g} = \frac{1}{2\pi}\Sigma_k g_k $ that has the following Keldysh form
\begin{eqnarray}
\bar{g}&=&\left(
         \begin{array}{cc}
              \bar{g}^R & \bar{g}^K    \\
               0 & \bar{g}^A     \\
               \end{array}
                \right)  \label{keld}
\end{eqnarray}
Here all entries are $4\times4$ diagonal matrix in Nambu space. In energy representation each of them has a form: $\bar{g}^{R,A}=\mp \frac{i}{2}N(0)$  diag(1,1,1,1) and
$i\bar{g}^K$= N(0)diag$[\tanh\frac{\omega-eV}{2T}(1,1,0,0)+\tanh\frac{\omega+eV}{2T}(0,0,1,1)]$, where N(0) is the electron density of states in the normal metal lead. To calculate the noise we need to find the effective action as function of a quantum source. We combined the current operator $(J)$ multiplied by source fields $\hat{\lambda}$=diag$(\lambda_1,-\lambda_2)$  on standard $1,2$ Keldysh contour. We obtain a combined source-tunneling contribution to the action
\begin{eqnarray}
  S_{s} &=& -\frac{1}{2}\int dt\sum_{i}[\bar{c}(0)(\tau_z \sigma_z + \hat{I}\hat{\lambda}) \hat{V}_i \gamma_i \nonumber\\
  &&+ \gamma_i\hat{V}_i^{\dagger}(\tau_z \sigma_z - \hat{I}\hat{\lambda}) c(0)]
\end{eqnarray}
Here the Pauli matrix $\sigma_z$ relates quantum operators  (Majorana $\gamma_i=(\gamma_{i1},\gamma_{i2})$ and lead fermions)  on the two time contours $(1,2)$. $\hat{I}$ is the unit matrix in the Keldysh or Nambu spaces.
Integrating out the lead electron operators and performing a rotation to the Keldysh form similar to (\ref{keld}) we arrive to a simple representation of $S_s$. Keeping only the quantum source $\lambda_q=(\lambda_1-\lambda_2)/2$ we obtain
\begin{eqnarray}
  S_{s} &=& -\pi \int dt\sum_{i,j}[ \gamma^{'}_i\hat{V}_i^{\dagger}(\tau_z \sigma_x -\nonumber\\ && \hat{I}\hat{I}\lambda_q) \bar{g} (\tau_z \hat{I} + \hat{I}\sigma_x\lambda_q)\hat{V}_j\gamma^{'}_j ] \label{se}
\end{eqnarray}
Where $\gamma^{'}_i=(\gamma^{cl}_i,\gamma^{q}_i)^T=(\gamma_{1i}+\gamma_{2i},\gamma_{1i}-\gamma_{2i})^T/2$.
We also notice a subtlety: due to the self-conjugacy condition the Keldysh GF for Majorana fermions differs from (\ref{keld})  and is similar to that of boson fields
\begin{eqnarray}
G&=&\left(
         \begin{array}{cc}
              G^K & G^R    \\
               G^A & 0     \label{mkel}\\
               \end{array}
                \right)
\end{eqnarray}
 To zero order in tunneling to the lead the action follows from Eq.(\ref{HM})
\begin{eqnarray}
S_M&=&\frac{1}{2}\int dt dt' [\gamma^{'}_i (t)]^{T}G^{-1}_{0ij}(tt')\gamma^{'}_j(t')
\end{eqnarray}
here the  inverse matrix GF is
\begin{eqnarray}
[G_0^R]^{-1}&=&\frac{1}{2}[i\partial_t-2i\hat{t}]
\end{eqnarray}
The total GF follows immediately if we add the zero source part contribution of $S_s$ to the $G_0^{-1}$:
\begin{equation}\label{G}
G^{-1}=G_0^{-1}-2\pi \hat{V}^{\dagger}\sigma_x \bar{g}\hat{V}
\end{equation}
which can be used to verify the form (\ref{mkel}). The effective action with the source term becomes
\begin{eqnarray}
S_{eff}&=&\frac{1}{2}\int dt \gamma^{'} (G^{-1}+ Q(\lambda_q))\gamma^{'}\\
Q(\lambda_q)&=&\hat{V}^{\dagger}[\tau_z( \lambda_q \bar{g}-\sigma_x\bar{g}\sigma_x\lambda_q) +\lambda_q\bar{g}\sigma_x\lambda_q]\hat{V}
\end{eqnarray}
The partition function for calculations of transport and noise is a function of quantum field $\lambda_q$:
 $Z(\lambda_q)=\int D\gamma \exp[iS_{eff}]$ and $\ln Z(\lambda_q)=Tr\ln[1+G Q(\lambda_q)]$.
Taking the variation of $ \ln Z$ on $\lambda_q\rightarrow 0$ we get expressions for the average current and the noise power.
\begin{eqnarray}
  I &=& \frac{e}{2\hbar} \frac{\delta \ln Z}{\delta \lambda_q}=\frac{e}{2h}\frac{\delta }{\delta \lambda_q}\int d\omega Tr[G Q] \nonumber\\
  S_n (tt')&=& -\frac{1}{2} \frac{\delta^2}{\delta \lambda_q(t)\delta \lambda_q(t')}\ln Z \label{Sn}
\end{eqnarray}

Before presenting explicit formulas
 we discuss the approximations which are used: (a)  We consider  the tunneling element ${\bar f}_{i\sigma}$ as spin independent. (b) When (\ref{G}) is inverted we neglect off diagonal ${\bar f}_{i\sigma}{\bar f}_{j\sigma'}$ terms that involve oscillation with a Fermi wavevector which vanish upon averaging on large spacings between MBS \cite{flens}.  (c) We take a wide band limit ignoring the momentum dependence of $V_i$. Thus we have $\Gamma_{ij}=4\pi N(0)\bar{f}_i^2\delta_{ij}$. (d) We also consider $\Delta >\Gamma{ij}$.
 We proceed now to take explicitly the trace in Keldysh and Nambu spaces leading to general expressions  for the current and the total zero frequency noise power $S_n=S_1+S_2$
\begin{eqnarray}
 I &=& \frac{e}{2h} \int d \omega Tr (\Gamma Im G^R)[\tanh\frac{\omega_-}{2T}-\tanh\frac{\omega_+}{2T}] \label{II} \\
  S_{1} &=& \frac{2e^2}{h} \int d \omega Tr (\Gamma Im G^R)[1-
\frac{1}{2}(\tanh^2 \frac{\omega_-}{2T}+ \tanh^2 \frac{\omega_+}{2T})]   \nonumber \\
 S_2&=& \frac{e^2}{h} \int d \omega Tr (\Gamma Re G^R\Gamma Re G^R)
(\tanh \frac{\omega_-}{2T}- \tanh \frac{\omega_+}{2T})^2 \nonumber
\end{eqnarray}
Here $\omega_{\pm}=\omega\pm eV$, $Im G^R=(G^R-G^A)/(2i)$ and $Re G^R=(G^R+G^A)/2$. The term $S_1$ mainly contributes to the thermal part of the noise power. In the limit of zero temperature ($T\rightarrow0)$ $S_1\rightarrow0$. The $S_2$ term for $V>T\rightarrow0$ defines the shot noise which is the total noise in this temperature limit
\begin{equation}\label{St}
   S_{shot} =\frac{8e^2}{h} \int_0^{eV} d \omega Tr (\Gamma Re G^R\Gamma Re G^R)
\end{equation}

{\it Isolated Majorana states}: For single Majorana state the matrix $\hat{t} $=0 and $G^R=G^R_{11}=2/(\omega+2i\Gamma)$. First let us consider the zero bias conductance in the limit $T>eV\rightarrow 0$. In this case $S_n$ corresponds to  thermal noise.  Directly by calculating the current (\ref{II}) or with the help of fluctuation dissipation theorem: $\frac{\partial S_n}{4\partial T}=\sigma$ we obtain the conductance
\begin{equation}\label{cm1}
    \sigma=\frac{2e^2}{h}\frac{4\Gamma^2}{T^2+4\Gamma^2}
\end{equation}
At zero temperature and $V\neq0$ the differential conductance coincides with Eq.(\ref{cm1}) if we replace $T$ by $eV$ \cite{flens}, while the shot noise for single Majorana state acquires the form
\begin{equation}\label{stm1}
  S_n=\frac{8e^2\Gamma}{h}(\arctan\frac{eV}{2\Gamma}-\frac{2eV\Gamma}{(eV)^2+4\Gamma^2})
\end{equation}
More complicated formulas  for $S_n$ follow when two MBS couple by a tunneling element $t$ and couplings $\Gamma_{11},\Gamma_{22}$ to the lead. In comparison, the well known
Andreev contribution to the noise of normal-superconductor (NS) junction \cite{blanter} is of order
 \begin{equation}\label{sns}
    S\sim\frac{4e^3 V}{h}(\frac{\Gamma}{\Delta})^2\,.
 \end{equation}
 This is much weaker then (\ref{stm1}) or the other MBS networks or by the factor $(\Gamma/\Delta)^2\ll 1$.

 Fig.1 presents the conductance and shot noise power as function $U=eV/\Gamma$ for different realizations. With two coupled Majorana's  the shot noise shows clear steps in its voltage dependence. They corresponds to the peaks in conductance at $eV=\pm 2 t$. Therefore the measurement of the noise can detect Majorana fermions. In case that one of the links is weak:  $t_{link}<t$ the noise loses its structure ( lower right panel in Fig.1) and becomes similar to the case of a single Majorana (upper left panel in Fig. 1). In this case  the sharp drop of conductance at voltage $eV\rightarrow 0$ can be smeared by a small temperature.

\begin{figure}
\begin{center}
\includegraphics [width=0.5 \textwidth ]{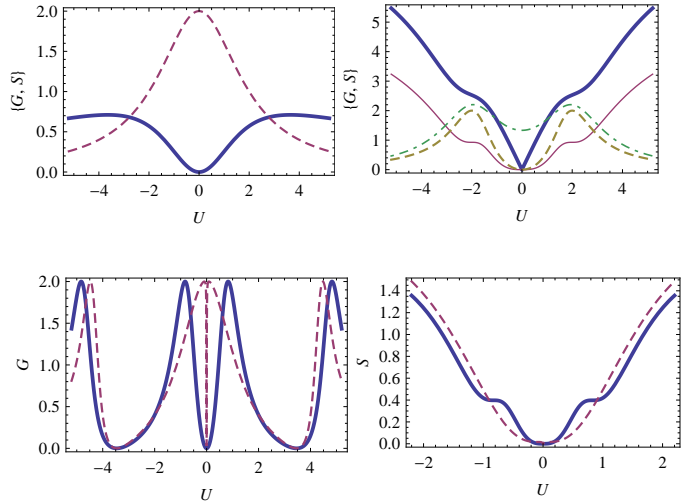}
\caption {(a) Upper pannels for conductance (dashed or dot-dashed line) and shot noise (full lines): Upper left for a single MB, upper right for two MBS with $t=\Gamma$ -- upper lines (dot dashed and thick lines) for $\Gamma_{11}=2\Gamma_{22}=\Gamma$ while lower lines (dashed and thin lines) for $\Gamma_{11}=\Gamma, \Gamma_{22}=0$. (b) Lower pannels  for four MBS: lower left conductance for weak $t_{12}=0.1\Gamma$ (dashed line) and strong $t_{12}=\Gamma$ (full line); all $t_{i,i+1}$ are equal and only the 1st MBS is coupled to the lead. Lower right has the corresponding shot noise. The conductance G is presented in  units $e^2/h$  and the shot noise power $S$ in units of $2\Gamma e^2/h$. }
\end{center}
\end{figure}
The Fano factor is defined as the ratio $F=S_{shot}/2eI$. For the NS junction $F(V\rightarrow 0)=2$ reflecting the transmission of cooper pairs through the superconductor. The Fano factor corresponding to the upper right panel of Fig. 1 (2 MBS) is plotted in Fig. 2.
The zero bias value of Fano factor $F(V=0)=2$ does not depend on the on the value of $t\neq0$ and is the same integer for four MBS, the case in the lower left panel of Fig. 1. However,
when the second Majorana state interacts with the normal lead, i.e. $\Gamma_{22}\neq 0$, $F(V=0)$ is non-integer and decreasing, as shown by the solid line in Fig. 2. In contrast, with odd number of MBS $F(V\rightarrow 0)=0$.
\begin{figure}
\begin{center}
\includegraphics [width=0.4 \textwidth ]{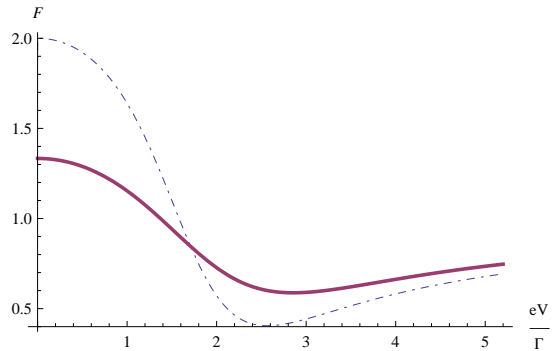}
\caption {Fano factor $F$ for a double MBS corresponding to the upper right panel of Fig.1. The dot-dashed curve shows $F(V)$ when only the 1st MBS is coupled to the lead, while the solid line stands for case when both MBS couple to the lead: $\Gamma_{11}=\Gamma$ and $\Gamma_{22}=\Gamma/2$. }
\end{center}
\end{figure}

 Considering next the case of an infinite homogeneous chain, where all near neighbor couplings are identical $t_{i,i+1}=t$, the Green's functions can be easily found \cite{flens}. The calculations of the shot noise are performed for $t/\Gamma=0.1$ and compared with conductance in Fig. 3.

\begin{figure}
\begin{center}
\includegraphics [width=0.4 \textwidth ]{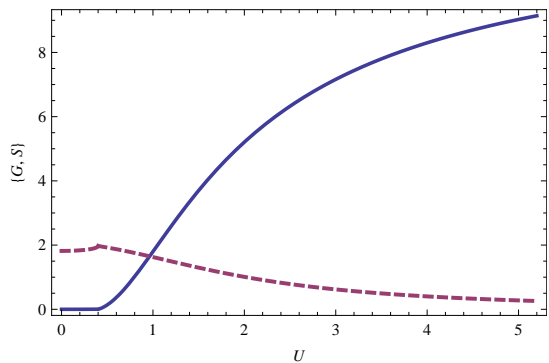}
\caption {Conductance (dashed curve) and shot noise (solid line) for tunneling into a homogeneous infinite Majorana chain as function of $U=eV/\Gamma$  for $t=0.1\Gamma$.}
\end{center}
\end{figure}

{\it Disordered case -- isolated Majorana pairs}:
We consider next the case of a semi infinite MBS chain with a random $t_{i,i+1}$ and only the 1st MBS is coupled to the lead $\Gamma_{ii}\sim \delta_{i,1}$. This system can be mapped to a system of independent MBS pairs \cite{refael} by using analysis methods of the quantum Ising spin chain \cite{yang,fisher}. The key ingredient is to isolate the strongest bond at each decimation. The MBS pairs have an an effective hopping
$\epsilon_n^{max}\approx \eexp{-\eexp{n/2}}$ (having a typical weight) leading to peaks of conductance at $eV=\epsilon_n$ with weights
$\Gamma_n^2\approx \eexp{-8\Gamma_0 e^{n/2}}$;  $\Gamma_0$ is a starting logarithmic flow parameter chosen here as $\Gamma_0=0.1$. For a given voltage the contribution to the noise is from pairs with $\epsilon_n^{max}>eV$, resulting in jumps at $\epsilon_n=eV$.
We estimate the averaged current and shot noise using our formulas (\ref{II},\ref{St}). The relevant frequencies $\omega\leq \epsilon_n ^{max}$ are selected by a Fermi function $f=1/(1+\exp[(\omega-\epsilon_n )/\delta]$ with small $\delta$ and we approximate $G=[\omega-eV+i\delta']^{-1}$ with a small $\delta'$. We perform calculations for $n\leq 6$. The shot noise power as a function of a double logarithmic voltage dependence is presented by Fig.3. $S_n$ and the average current $J$ show similar behavior, though, the jumps are of different heights.
\begin{figure}
\begin{center}
\includegraphics [width=0.4 \textwidth ]{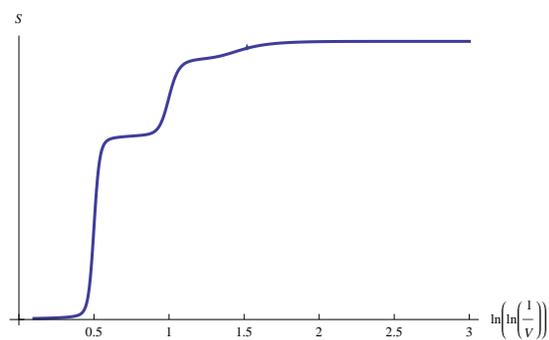}
\caption {The shot noise power as function of a double logarithmic voltage. The behavior of the noise power has similar sharp variations as the current \cite{refael}. }
\end{center}
\end{figure}

{\it Conclusion}:
 We apply the standard Keldysh technique  to calculate the
shot noise power of  a Majorana chain interacting with a normal lead. We find that the shot noise has markedly distinct forms for different numbers of MBS and for long ordered or disordered chains. We show that for weak coupling to the lead the shot noise as well as the conductance are much stronger then the corresponding SN junction. We calculate the Fano factor and showed that its value crucially depends on whether  even or odd number of MBS are involved. The Fano factor is similar in magnitude to the SN case, in particular $F(V\rightarrow 0)=2$ for even number of MBS and if only one MBS is coupled to the lead, as for SN junction; however, when more than one MBS is coupled to the lead (Fig. 2) $F(0)$ is non-integer.

\begin{acknowledgments}
We would like to thank E. Grosfeld  and A. W. W. Ludwig for stimulating discussions.
This research was supported by THE ISRAEL SCIENCE FOUNDATION (grant No. 1078/07).
\end{acknowledgments}

\end{document}